\documentclass[9pt,twocolumn,twoside]{opticajnl}
\journal{opticajournal} 

\setboolean{shortarticle}{true}

\newcommand{\vvss}{``}
\newcommand{\vvdd}{''}
\newcommand{\vv}[1]{\vvss #1\vvdd}
\usepackage{amsmath} 
\usepackage{lineno}
\usepackage{xcolor}
\usepackage{soul}
\usepackage{ulem}
\usepackage{lineno}

\title{Cavity ring-down spectroscopy at 2-$\mu$m wavelength assisted by a comb-locked optical parametric oscillator}

\author[1]{Vittorio D'Agostino}
\author[1]{Eugenio Fasci}
\author[1]{Muhammad Asad Khan}
\author[1]{Stefania Gravina}
\author[1]{Livio Gianfrani}
\author[1,*]{Antonio Castrillo}

\affil[1]{Department of Mathematics and Physics, Universit\`{a} degli Studi della Campania \vv{Luigi Vanvitelli}, 81100, Caserta, Italy}

\affil[*]{antonio.castrillo@unicampania.it}

\begin{abstract}
We report on a comb-locked cavity ring-down spectrometer developed for high-precision molecular spectroscopy at 2 $\mu$m. It is based on the use of an external-cavity diode laser that is offset-frequency locked to the signal output of a singly-resonant optical parametric oscillator. This latter acts as reference laser, being locked to a self-referenced optical frequency comb, which in turn is stabilized against a GPS-disciplined Rb-clock.
The performance of the spectrometer is investigated by probing a pair of N$_2$O transitions belonging to hot vibrational bands. One of these, never observed before, is included in the N$_2$O line list of the ExoMol database. Absolute center frequencies are retrieved with a 1-$\sigma$ global uncertainty of 108 kHz.
\end{abstract}

\setboolean{displaycopyright}{false} 

\begin{document}

\maketitle

Remote sensing of greenhouse-relevant molecules (CO$_2$, CH$_4$ and N$_2$O, to cite a few examples) is based on physical models that need spectroscopic parameters (line intensity, collisional broadening and shifting parameters), which are usually collected in databases like HITRAN \cite{HITRAN}, GEISA \cite{GEISA}, and CDSD-296 \cite{CDSD-296}. Unfortunately, these data are usually provided with a relative accuracy of about 1\% at best. In the case of atmospheric CO$_2$, current and upcoming missions, including the NASA's Orbiting Carbon Observatory (OCO-2 and OCO-3) \cite{Thompson2012}, the JAXA's Greenhouse gases Observing SATellite (GOSAT-1 and GOSAT-2) \cite{Imasu2023}, and the CNES's MicroCarb \cite{MicroCarb}, use satellite-based spectrometers that require high-quality parameters at 1.6 and 2.06 $\mu$m. In the case of N$_2$O, the Infrared Atmospheric Sounding Interferometer (IASI) \cite{IASI}, onboard the EUMETSAT Metop satellite, and the TROPOspheric Monitoring Instrument (TROPOMI) \cite{Tropomi}, on board the Copernicus Sentinel-5 Precursor satellite, make use of a similar approach for the determination of global-scale N$_2$O sources. For both molecules, the achievement of the ambitious target of a relative uncertainty of 0.25\% in the determination of their mole fractions requires that line intensities, collisional broadening and shifting parameters are known with higher accuracy \cite{Thompson2012}. The demand of such data has motivated several groups in developing experimental setups that could guarantee suitable levels of accuracy. In the case of CO$_2$, transition frequencies have been determined with a 10$^{-12}$ uncertainty level in the 1.57 \cite{Reed2020} and 2.06 $\mu$m \cite{Fleurbaey2023} spectral regions. As for line intensities, the accuracy has been pushed well below the 1\% level at the 2 $\mu$m wavelength \cite{FLEURBAEY2020,Odintsova2017}, even achieving the sub-promille level at 1.6 $\mu$m \cite{Fleisher2019}. Similar efforts have been performed for N$_2$O line intensities \cite{Odintsova2020,Adkins2022}. It is important to note that these studies have benefited from the superior performance guaranteed by cavity-enhanced techniques, such as cavity ring-down spectroscopy (CRDS) and optical feedback cavity-enhanced absorption spectroscopy (OFCEAS), often assisted by optical frequency comb (OFC) synthesizers \cite{Gianfrani2024}. In this regard, comb referencing of the probe laser, while preserving the tunability of the laser itself, is one of the key ingredients that allows one to achieve high accuracy in conjunction to traceability to the International System (SI) of Units. The possibility of satisfying these requirements around 2 $\mu$m would be relevant not only for atmospheric monitoring applications but also for testing quantum chemistry calculations for linear molecules such as CO$_2$ \cite{Polyansky2015,Zak2017} and N$_2$O \cite{Tennyson2024,Yurchenko2024}.

Here, we report on a comb-assisted CRDS spectrometer, operating in the 2 $\mu$m wavelength range, based on the use of a widely tunable external-cavity diode laser (ECDL) which is frequency-locked to the signal output of a singly-resonant optical parametric oscillator (OPO). The OPO signal acts as a reference laser, being weakly locked to the nearest tooth of a self-referenced OFC, which, in turn, is phase-locked to a GPS-disciplined Rb-clock. This is a new type of usage for an OPO source. In fact, to the best of our knowledge, the idler radiation is typically employed for highly precise investigations of molecular spectra in the mid-infrared region \cite{Zhang2020-1,Zhang2020-2,Zhong2025,Vainio2016, Ricciardi2012}. To investigate the performance of the CRDS spectrometer, we recorded Doppler-limited absorption spectra of N$_2$O as a function of the gas pressure, thus retrieving the zero-pressure transition frequency and the pressure-induced broadening and shifting coefficients for a pair of vibro-rotational lines. 

\begin{figure}[ht]
\centering
\includegraphics[width=\linewidth]{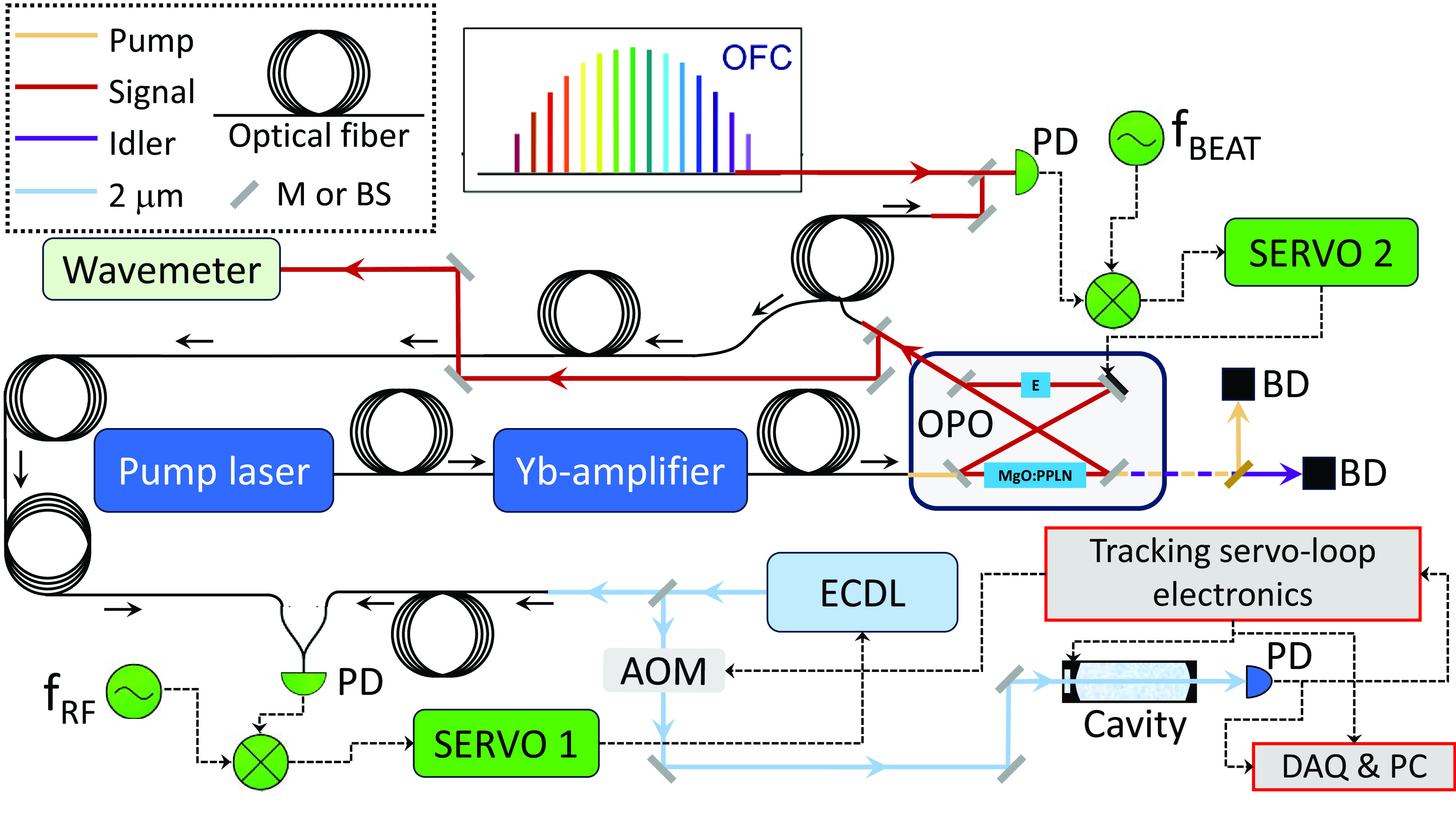}
\caption{Schematic of the experimental setup. OFC stands for optical frequency comb; ECDL, external-cavity diode laser; OPO, optical parametric oscillator; E, etalon; PD, photodiode; AOM, acoustic-optic modulator; BD, laser beam dump; DAQ, acquisition board; PC, personal computer. Black arrows indicate the direction of propagation of the laser beams in the optical fibers.}
\label{fig:fig1}
\end{figure}

The spectrometer is schematically reported in Figure \ref{fig:fig1}. It includes an OFC, a 2-$\mu$m ECDL, a singly-resonant continuous-wave (cw) OPO source, and a high-finesse optical cavity.  
The OFC is a commercially available self-referenced optical frequency comb, based on an erbium-doped fiber laser. Its repetition
rate ($f_{REP}$=250 MHz) and carrier-envelope offset frequency ($f_{CEO}$=20 MHz) are stabilized against the frequency of a GPS-disciplined Rb-clock. This latter has a relative drift of the time signal over
1 week of $\sim$3$\times$10$^{-14}$ (Allan deviation when locked to the GPS satellites).The OFC provides a supercontinuum in the wavelength
range between 1050 and 2100 nm. The absolute frequency of each tooth, $f_{N}$, is given by the well-known equation $f_N=N\times f_{REP}+f_{CEO}$, $N$ being the mode order. 

The ECDL is mounted in the so-called Litmann-Metcalf configuration.  It emits 20 mW in the 1980-2100 nm range, its full-width half-maximum (FWHM) being about 200 kHz at 1 ms of integration time. It is offset-frequency locked to the signal output of the OPO source following the scheme reported in Ref. \cite{Castrillo2010}. Briefly, a phase/frequency detector compares the signal of a RF synthesizer ($f_{RF}$) with the 80-times scaled frequency of the ECDL-OPO beat note, the latter being detected by a fiber-coupled 12.5-GHz bandwidth InGaAs photodetector. The error signal produced by the phase/frequency detector is properly integrated by a PID servo, which provides a correction signal to the laser external cavity piezo and to the injection current. In this way a constant frequency offset between the ECDL and the OPO signal can be maintained. The bandwidth of the entire control loop has been determined to be about 10 kHz. Thus, a tuning of $f_{RF}$ results in a continuous and highly-accurate scan of the ECDL, which can be as wide as 10 GHz. The ECDL frequency, $f_{ECDL}$, can be easily determined using the equation $f_{ECDL}=f_{OPO}\pm f_{RF}$, $f_{OPO}$ being the frequency of the OPO signal. The $\pm$ ambiguity can be solved once it is known whether the ECDL frequency is red- or blue-shifted compared to $f_{OPO}$.

The singly-resonant OPO is pumped by a distributed-feedback diode laser, emitting at 1064 nm and amplified by a Yb-doped fiber amplifier up to 10 W. It is based on a bow-tie ring cavity configuration, the cavity mirrors being highly reflective in the range from 1.4 to 2.07 $\mu$m (possible wavelengths of the signal) and transparent at the pump and idler wavelengths. A temperature-stabilized non-linear MgO:PPLN crystal is placed inside the cavity. It can be continuously translated for a coarse tuning of the wavelength. An etalon is placed in the secondary focus of the cavity so that it is possible to perform a stable single-longitudinal-mode selection, while one of the cavity mirrors is equipped with a piezo for fine tuning. The crystal temperature and position, the etalon angle, and the cavity length directly affect the frequency of the signal, while the idler frequency is determined by energy conservation. The average output power of the signal and idler are about 2 and 1 W, respectively. It is worth noting that, for the purposes of the present work, the idler beam is not used, while the signal is employed as reference laser for the experiment. In fact, once the desired signal frequency is reached, its long-term stabilization is accomplished by means of the use of the OFC. Thus, the beat note between the signal and the nearest tooth of the OFC, $f_{BEAT}$, is properly processed to generate a correction signal that actively controls the length of the OPO cavity through the piezo, with a bandwidth of few tens of Hz, which enables us to correct for possible frequency drifts of the cavity during the spectral acquisitions. This expedient also ensures the absolute determination of $f_{OPO}$, which is equal to $N\times f_{REP}\pm f_{CEO}\pm f_{BEAT}$. The $\pm$ signs are easily determined by slightly varying $f_{REP}$ and $f_{CEO}$ and observing the variation of $f_{BEAT}$, whereas the tooth order is obtained by measuring the signal wavelength by means of a wavemeter having a 30-MHz accuracy. During the experiment, we lock $f_{BEAT}$ to 20 MHz, while N=599353. It is useful to remind that the mode-hop-free tuning range of the signal is limited to few tens of MHz. 

The high-finesse cavity consists of two plano-concave high-reflectivity mirrors, spaced 43 cm apart by a Zerodur block. The mirrors have a radius of curvature of 1 m and a nominal reflectivity greater than 99.99\%. One of the mirrors is equipped with a piezo transducer for a fine tuning of the cavity length. The ring-down time under vacuum conditions is about 24 $\mu$s. At the output of the cavity, a 1 MHz InGaAs detector is used to monitor the light emerging from the resonator. A digital acquisition board (DAQ) records the ring-down events. The DAQ works at a sample rate of 2$\times$10$^6$ samples s$^{-1}$, its vertical resolution being 16 bit. The ECDL is sent into an optical isolator, followed by a AOM, whereas the first-order diffracted beam is coupled to the high-finesse cavity by means of a mode-matching telescope. The frequency of the laser light emerging from the AOM is up-shifted by a constant amount $f_{AOM}$=80 MHz. The AOM is also used as an optical switch to initiate ring-down decays. As reported in Refs. \cite{Fasci2018,Castrillo2024}, a tracking servo loop provides trigger signals and allows for fast acquisition of ring-down signals. The cavity is equipped with a calibrated pt-100 thermometer and a 100 Torr full-scale absolute pressure gauge for temperature and pressure readings, respectively. The relative accuracy of gas pressure measurements amounts to 0.05\%. 

Once a ring-down event is recorded, it is fitted to an exponential decay curve, yielding the decay time $\tau$. The absorption coefficient, $\alpha$, as a function of the laser frequency $\nu$ (in cm$^{-1}$), is determined from the well-known equation $\alpha(\nu)=\frac{1}{c}\Big(\frac{1}{\tau(\nu)}-\frac{1}{\tau_0}\Big)$, $c$ being the speed of light (in cm/s) and $\tau_0$ the decay time under vacuum conditions. It is worth noting that the probe laser frequency is given by $f_{ECDL}+f_{AOM}$, namely $N \times f_{REP} \pm f_{CEO} \pm f_{BEAT} \pm f_{RF} + f_{AOM}$.

\begin{figure}[ht]
\centering
\includegraphics[width=\linewidth]{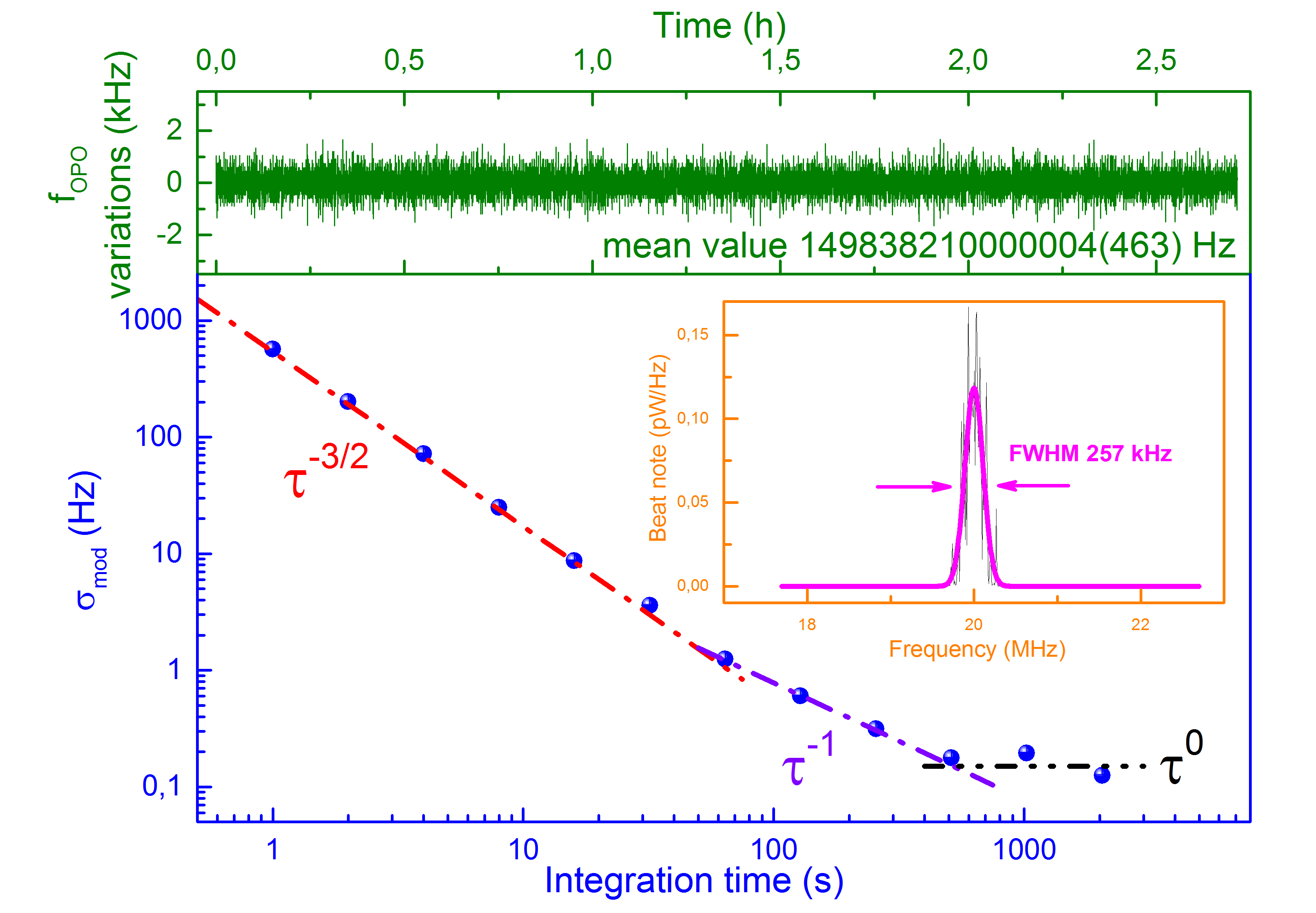}
\caption{Upper panel: frequency fluctuations
of $f_{OPO}$. Lower panel: modified Allan deviation as a function of the integration time for the time series reported in the upper panel. Inset: example of a beat-note between the OPO signal and the OFC.}
\label{fig:fig2}
\end{figure}

The beat note between the OPO signal and the nearest comb tooth is shown in the inset of Figure \ref{fig:fig2}. Its FWHM, as determined by fitting to a Gaussian function, is found to be 257 kHz. This measurement gives an estimation of the signal linewidth. In fact, considering that each comb tooth has a width of $\sim$110 kHz \cite{Gravina2024}, the FWHM of the OPO signal turns out to be about 230 kHz.

The upper panel of Figure \ref{fig:fig2} shows the fluctuations of $f_{OPO}$ over a period of about 3 hours, demonstrating the absence of any drifts. The in-loop relative stability of the signal frequency, as determined by the modified Allan analysis reported in the lower panel of Figure \ref{fig:fig2}, is 3.8$\times$10$^{-12}$ for an integration time of 1 s. This analysis reveals that a trend that is typical of a white phase-noise is observed for integration time between 1 and 70 s. It also indicates that from 70 to 500 s a Flicker phase-noise becomes prevalent, while after 500 s the contribution to the noise takes the behavior of a Flicker frequency-noise, the corresponding level being 0.15 Hz. We can conclude that the long-term stability of $f_{OPO}$ is limited by the stability of the Rb-clock.

\begin{figure}[ht]
\centering
\includegraphics[width=\linewidth]{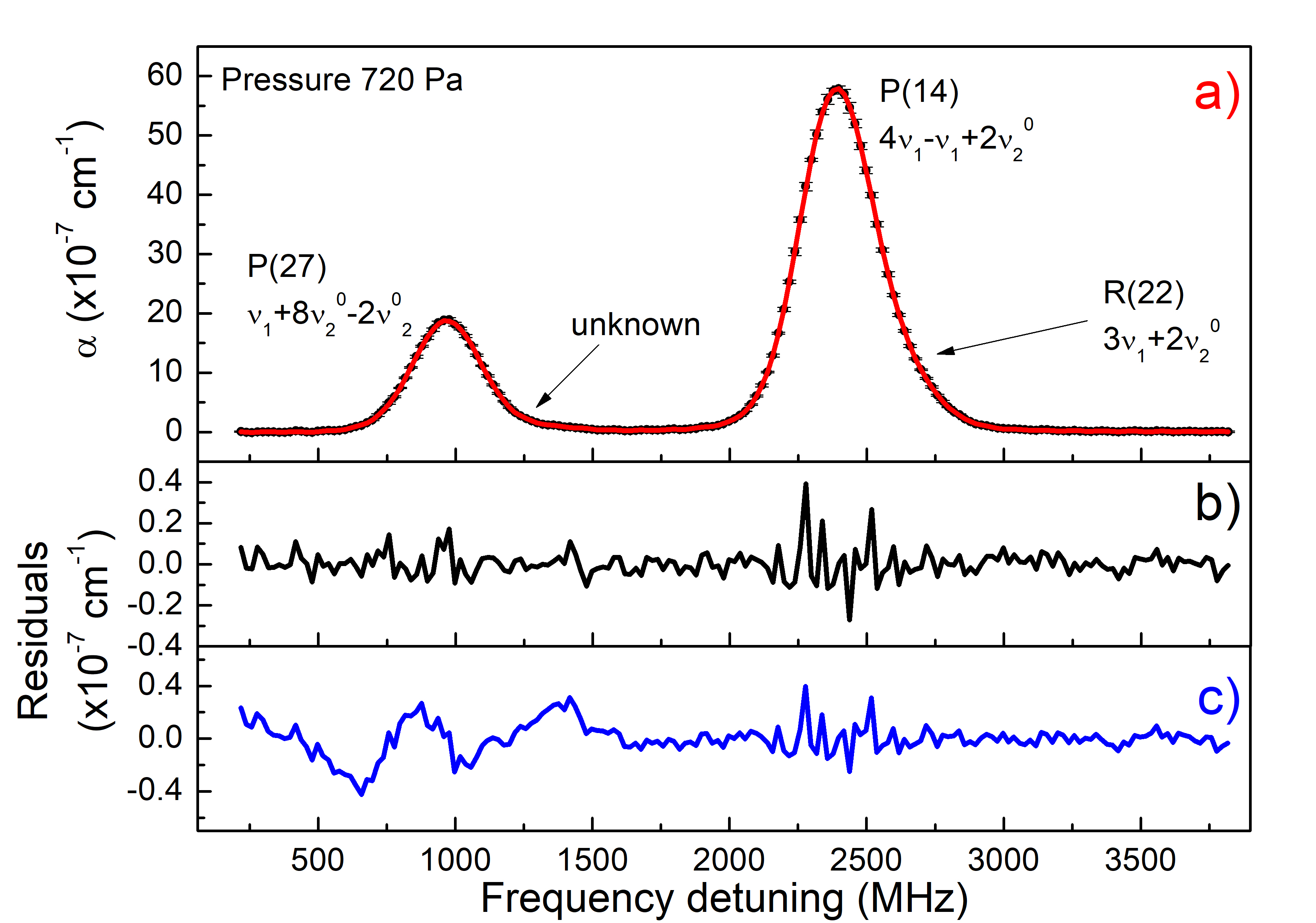}
\caption{Panel a): example of CRDS N$_2$O spectrum at a temperature of 296.4 K. Panel b): example of residuals, as obtained from the application of the fitting procedure. Panel c): as panel b) without the inclusion of the unknown transition. The increase of the residuals, in coincidence with the peak of the P(14) transition, may be ascribed to a significant decrease of the ring-down time.}
\label{fig:fig3}
\end{figure}

To test the performance of the CRDS setup, we recorded absorption spectra of N$_2$O at $\sim$5000 cm$^{-1}$. The upper panel of Figure \ref{fig:fig3} shows a 3.6-GHz wide nitrous oxide spectrum, averaged over 10 consecutive scans, at a pressure of 720 Pa (of a 99.0\% pure N$_2$O gas sample), obtained in about 5 minutes. This spectrum consists of 181 spectral points, equally spaced of 20 MHz. According to the HITRAN database \cite{HITRAN}, the main absorption line is associated to the P(14) transition of the $4\nu_1-\nu_1+2\nu_2^0$ hot band at 4997.979680 cm$^{-1}$, while the smaller line on its right wing can be ascribed to the N$_2$$^{18}$O R(22) line (of the $3\nu_1+2\nu_2^0$ band) at 4997.980988 cm$^{-1}$ \cite{HITRAN}. The absorption peak at a lower frequency cannot be found in HITRAN, whereas it is included in ExoMol, a useful database that contains theoretical line lists for a wide variety of molecules of interest for astrophysical environments such as exoplanets, brown dwarfs, and cool stars \cite{Tennyson2016}. This database has recently been updated with a new N$_2$O line list \cite{Yurchenko2024}. Here, a transition at 4997.932385 cm$^{-1}$ has been reported and assigned to the P(27) component of the $\nu_1+8\nu_2^0-2\nu_2^0$ hot band. According to ExoMol, its intensity should be 3.3 lower compared to the P(14) line. For the latter, a frequency of 4997.980516 cm$^{-1}$ is reported \cite{Yurchenko2024}, thus showing a positive shift of about 25 MHz compared to HITRAN. Information has not been found for the transition specified as unknown in Figure \ref{fig:fig3}. Its presence is clear if line fitting is repeated without considering it (see the residuals in the panel c) of Figure \ref{fig:fig3}). This is probably due to another molecule among the residual impurities of the gaseous sample. 

The spectrum of the upper panel of Figure \ref{fig:fig3} is fitted by means of the sum of a pair of symmetric speed-dependent Voigt profiles, as derived from the Hartmann-Tran profile (HTP) \cite{Ngo2013} setting to zero the velocity-dependent shifting-coefficient, the effective frequency of the velocity-changing collisions, and the partial correlation parameter. It must be noted that the two low-intensity interfering lines, namely, the R(22) and the unknown component, are both included in the spectra analysis, being modeled with a pair of Voigt convolutions. In the fitting procedure, the Doppler widths were constrained to be the same for all N$_2$O profiles. Panel b) of Figure \ref{fig:fig3} shows the residuals of the fit, whose root-mean-square (rms) value results to be 6$\times$10$^{-9}$ cm$^{-1}$, limited only by the experimental noise. CRDS spectra have been acquired at seven different gas pressures, in the range 250-1000 Pa, and analyzed according to the procedure described above. Similar results have been found across the different pressures.

\begin{figure}[ht]
\centering
\includegraphics[width=\linewidth]{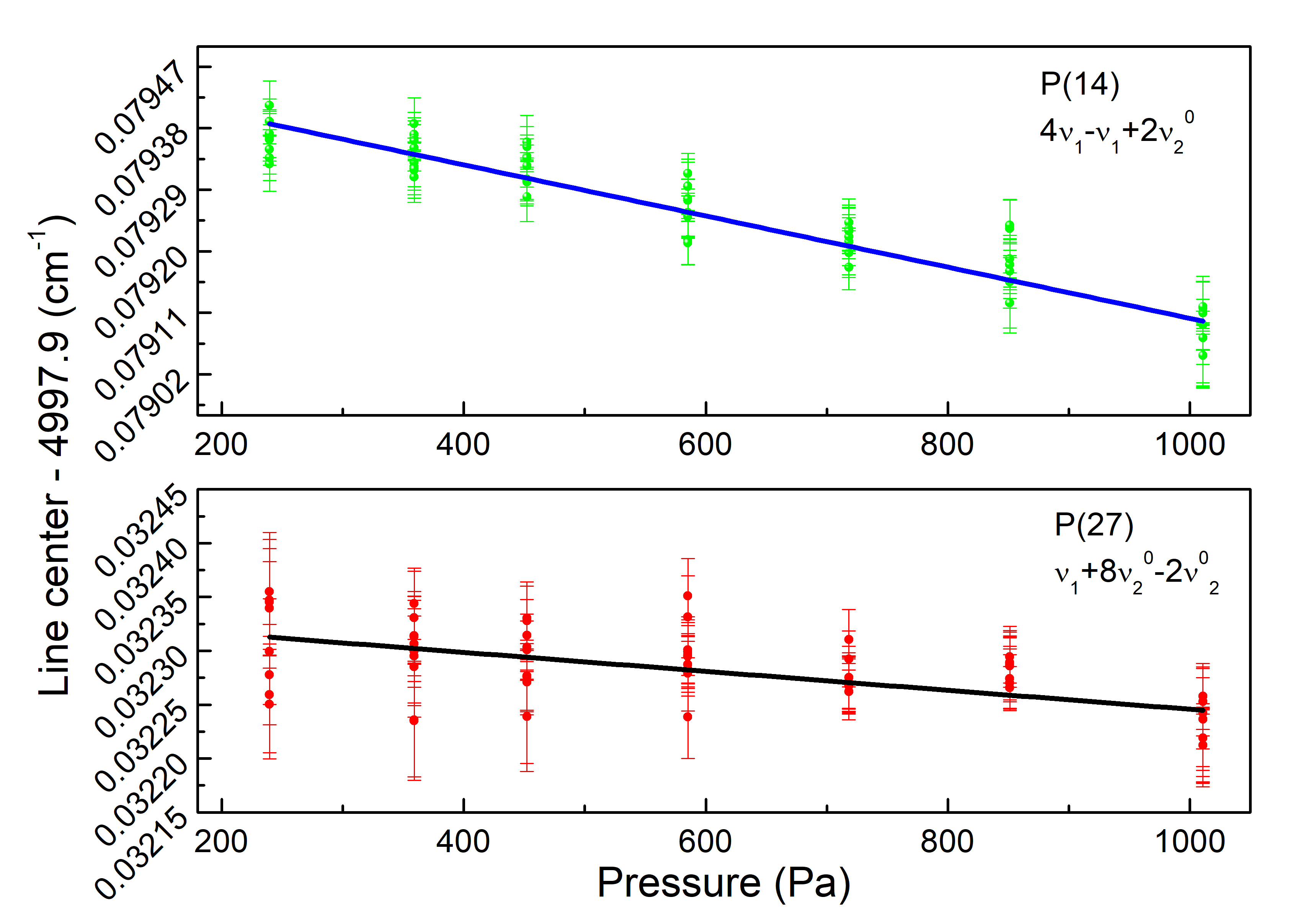}
\caption{Center frequencies retrieved as a function of the N$_2$O pressure. The weighted best-fit line allows one to determine the zero-pressure center frequency. Upper panel refers to the P(14) transition, while the lower one concerns the P(27) line.}
\label{fig:fig4}
\end{figure}

As a consequence of the fitting procedure, the line center frequency as a function of the pressure could be determined for the two transitions. The results of this analysis are summarized in the upper and lower panels of Figure \ref{fig:fig4} for the P(14) and the P(27) transition, respectively. From a weighted linear fit of these data, it is possible to extract the zero-pressure frequencies. In particular, for the P(14) line, a line center frequency of 4997.979472 cm$^{-1}$ is determined, the statistical uncertainty being 2$\times$10$^{-6}$ cm$^{-1}$. This value is only 6.2 MHz lower than the one reported in Ref. \cite{HITRAN}, thus demonstrating a relatively good agreement, if compared with the 3-MHz quoted uncertainty of HITRAN. The linear fit also provides an estimation of the pressure-induced self-shifting coefficients that is -3.7(8)$\times$10$^{-7}$ cm$^{-1}$/Pa. As for the P(27) line, a shifting coefficient of -9(7)$\times$10$^{-8}$ cm$^{-1}$/Pa and a line center frequency of 4997.932334(2) cm$^{-1}$ are determined. The latter is 1.5 MHz lower with respect to the ExoMol value, while a blue-shift of about 31 MHz is found for the P(14) line \cite{Yurchenko2024}. The same analysis allowed us to retrieve the self-broadening coefficient, equal to 0.122(3) and 0.073(5) cm$^{-1}$/atm for the P(14) and P(27) transition, respectively.

As for the uncertainty budget of the absolute frequency measurements, useful details are given below. The statistical contribution comes from the weighted linear fit of Figure \ref{fig:fig4}. The OFC contributes with an uncertainty of 1.7$\times$10$^{-8}$ cm$^{-1}$, which is due to the stability of the GPS-disciplined Rb clock. The uncertainty in the driving frequency of the AOM and the contributions originating from the wave front curvature can be neglected. Similarly, the second-order Doppler shift can be neglected, while the recoil shift is estimated to be 3.3$\times$10$^{-8}$ cm$^{-1}$.
Applying the procedure described in Ref. \cite{Castrillo2023}, we estimate that the gas pressure contributes with a systematic uncertainty of about 3$\times$10$^{-6}$ cm$^{-1}$. Varying the intra-cavity power, we did not observe any measurable influence in the line center frequencies within the experimental noise. Therefore, we can conclude that the global uncertainty of our frequency determinations is mostly due to the statistical contribution and to the pressure reading, the global uncertainty being comparable (or even better, in some cases) to similar Doppler-limited CRDS experiments \cite{Gatti2015}.

In conclusion, we developed a 2-$\mu$m wavelength CRDS spectrometer assisted by a comb-locked cw singly-resonant OPO. This latter acts as reference laser for the spectrometer, being stabilized against the nearest tooth of an optical frequency comb. As demonstration of the successful operation of the spectrometer, we recorded Doppler-limited N$_2$O spectra, as a function of the gas pressure, for a pair of spectral components of vibrational hot bands around 5000 cm$^{-1}$. Absolute center frequencies have been determined with an overall uncertainty of 108 kHz. Pressure-broadening and shifting coefficients have been also provided. In reason of the large tunability range of both the ECDL and the OPO signal (up to 90 nm), the spectrometer is well suited for precision measurements of spectroscopic parameters for a variety of atmospheric-relevant molecules, such as CO$_2$, in different wavelength windows in the near-infrared. More particularly, we plan to perform new stringent tests of \textit{ab-initio} quantum chemistry calculations of the line intensities for the CO$_2$ absorption bands in the 2-$\mu$m region, thus improving the results obtained in \cite{Odintsova2017}. Finally, a robust, widely tunable, comb-locked laser source at 2-$\mu$m wavelength might be of interest for a new generation of gravitational waves interferometers that are likely to shift their operation wavelength from the 1 to the 2 $\mu$m region \cite{Kapasi2020}.

\begin{backmatter}
\bmsection{Funding} This work was supported by the EURAMET Metrology Partnership project \vv{PriSpecTemp} [grant number 22IEM03]. This project has received funding from Metrology Partnership program co-financed
by the Participating States and the European Union’s Horizon 2020 research and motivation program.

\bmsection{Acknowledgment} 
SG is grateful to the Italian Ministry for University and Research for providing a Research Fellowship through the ETIC Project within the PNRR program. 

\bmsection{Disclosures} The authors declare no conflicts of interest.

\bmsection{Data availability} Data underlying the results presented in this paper are not publicly available at this time but may be obtained from the authors upon reasonable request.
\end{backmatter}


\end{document}